\documentclass[a4paper,12pt]{article}

\usepackage[dvips]{graphicx}
\usepackage{geometry}
\usepackage[centertags]{amsmath}
\usepackage{amsfonts}
\usepackage{amssymb}
\usepackage{amsthm}
\usepackage{latexsym}
\usepackage{fancyhdr}

\RequirePackage{hyperref}

\newcommand{\uD}{\ensuremath{\mathrm{d}}}
\newcommand{\uE}{\ensuremath{\mathrm{e}}}
\newcommand{\uI}{\ensuremath{\mathrm{i}}}

\renewcommand{\vec}[1]{\ensuremath{\boldsymbol{\mathrm{#1}}}}
\newcommand{\Cdot}{\ensuremath{\boldsymbol{\cdot}}}

\newcommand{\ArXivNo}{\href{http://arxiv.org/abs/quant-ph/0412103}{quant-ph/0412103}}

\newcommand{\XXXSize}{{\fontsize{12}{12}\selectfont\fbox{\textbf{\ArXivNo}}}}
\newcommand{\XXXTitle}{\hfill\XXXSize\newline\vskip 0.4cm}

\begin{document}
\pagestyle{fancy}

\sloppy

\title{\XXXTitle\textbf{Direct Derivation of the Schwinger Quantum
Correction to the Thomas--Fermi Atom}\thanks{Published in
\textit{International Journal of Theoretical Physics},
Vol.~\textbf{38}, No.~3, (1999) pp.~897--899. \ %
[\url{http://dx.doi.org/10.1023/A:1026613203875}] %
}}

\author{\textsc{Edouard~B.~Manoukian}\thanks{E-mail: \texttt{edouard@sut.ac.th}} \ and
\ \textsc{Parut~Bantitadawit} \\
{School of Physics, \ Suranaree University of Technology} \\
\ Nakhon Ratchasima, 30000, Thailand }
\date{} \maketitle

\begin{abstract}
The Schwinger quantum correction to the classic Thomas--Fermi atom
is directly derived by solving for the latter without recourse to
a modeling after the harmonic oscillator potential and without solving
for the particle density\@. \\
\end{abstract}

In an ingenious treatment of the quantum correction to the
remarkable Thomas--Fermi atom (Thomas, 1927~\cite{Thomas_1927};
Fermi, 1927~\cite{Fermi_1927}, 1928~\cite{Fermi_1928}), Schwinger
(1981)~\cite{Schwinger_1981} modeled his analysis after the
harmonic oscillator potential\@. \  Although this modeling
argument turns out to be correct, the importance of this ``atom,''
which has captivated physicists since its birth over 70 years ago
when quantum mechanics was still in its infancy, and will continue
to do so due to its extreme simplicity and remarkable success, has
motivated us to supply a direct derivation of the Schwinger
correction without recourse to a harmonic oscillator potential
modeling and without solving for the particle density (Dreizler
and Gross, 1990~\cite{Dreizler_1990})\@. \  The latter reference
also gives a fairly recent review of the state of the art of the
theory and gives extensive references to the monumental work of
Schwinger and to many other contributors. For more recent work and
additional references see Morgan (1996)~\cite{Morgan_1996}\@. \\

The quantum correction to the ground-state energy is given by the
compact expression
\begin{equation}\label{Eqn01}
  \delta{}E_{\mathrm{Qua}} = \int\!\!\uD^{3}\vec{r}\:\frac{2}{2\pi\uI}
  \int_{-\infty}^{\infty}\!\frac{\uD\tau}{\tau-\uI\varepsilon}\;\uI
  \frac{\partial}{\partial\tau}\left[
  \delta{}G_{0}(\vec{r}\tau,\vec{r}0;V_{\mathrm{TF}})
  -\delta{}G_{0}(\vec{r}\tau,\vec{r}0;V_{\mathrm{C}})\Big.\right]
\end{equation}
where the $\tau$-integral projects out the negative spectrum; \
$G_{0}(\vec{r}\tau,\vec{r}'0;V)$ is defined in terms of the Green
function:
$G_{\pm}(\vec{r}t,\vec{r}'0)=\mp(\uI/\hbar)\Theta(\mp{}t)
G_{0}(\vec{r}\tau,\vec{r}'0;V)$ with appropriate boundary
conditions $G_{\pm}(\vec{r}t,\vec{r}'0)=0$ for $t>0$ and $t<0$,
respectively\@. \  Here $\tau=t/\hbar$\@. \  $G_{\pm}$ satisfies
the differential equation
\begin{equation}\label{Eqn02}
  \left[-\uI\frac{\partial}{\partial\tau}-\frac{\hbar^2}{2m}\vec{\nabla}^{2}
  +V(\vec{r})\right]G_{\pm}(\vec{r}t,\vec{r}'0) = \delta^{3}(\vec{r}-\vec{r}')
  \delta(t).
\end{equation}
The potentials have the following familiar expressions~:
\begin{align}
  V_{\mathrm{TF}}(\vec{r}) &= -\frac{Ze^2}{r}\,f(r) \label{Eqn03} \\
  V_{\mathrm{C}}(\vec{r}) &=
  -\frac{Ze^2}{r}\left[1+f'(0)\frac{r}{a}\right] \label{Eqn04}
\end{align}
where $a=(3\pi/4)^{2/3}(\hbar/2me^{2})Z^{-1/3}$, $x=r/a$, and
$f(x)$ is the Thomas--Fermi function: $f(0)=1$, and vanishes like
$x^{-3}$ for $x\to\infty$\@. \   In Eq.~(\ref{Eqn01}) the
Coulombic contribution (\ref{Eqn04}), describing the so-called
tightly bound electrons near the nucleus, is appropriately
subtracted out\@. \   $\delta{}G_{0}$ denotes the shift from the
semi-classical limit\@. \\

Upon setting
\begin{equation}\label{Eqn05}
  G_{0}(\vec{r}\tau,\vec{r}'0;V) = \int\!\!\frac{\uD^{3}\vec{p}}{(2\pi\hbar)^3}
  \;\uE^{\uI\vec{p}\Cdot(\vec{r}-\vec{r}')/\hbar}\exp\left[-\uI\left(
  \frac{\vec{p}^2}{2m}\tau+U\right)\right]
\end{equation}
we have that $U$ satisfies the differential equation
($U\big|_{\tau=0}=0$)~:
\begin{equation}\label{Eqn06}
  -\frac{\partial}{\partial\tau}U+V(\vec{r})-\frac{\hbar}{m}\vec{p}\Cdot
  \vec{\nabla}U+\frac{\hbar^2}{2m}(\vec{\nabla}U)^{2}+\frac{\uI\hbar^{2}}{2m}
  \nabla^{2}U = 0.
\end{equation}
The semiclassical limit is given by $U_{0}=V(\vec{r})\tau$\@. \ It
is easily checked from (\ref{Eqn06}) that the leading shift
$\delta{}U$ in $U$ from the semiclassical limit is given by
\begin{equation}\label{Eqn07}
  \delta{}U = -\frac{\hbar\tau^2}{2m}\vec{p}\Cdot\vec{\nabla}V
  +\frac{\hbar^{2}\tau^{3}}{6m^{2}}(\vec{p}\Cdot\vec{\nabla})^{2}V
  +\frac{\hbar^{2}\tau^{3}}{6m^{2}}(\vec{\nabla}V)^{2}
  +\frac{\uI\hbar^{2}\tau^{2}}{4m}\nabla^{2}V.
\end{equation}
Upon replacing $U=U_{0}+\delta{}U$ in (\ref{Eqn05}) and carrying
out an elementary Gaussian integral, we obtain for $\delta{}G_{0}$
\begin{align}\label{Eqn08}
  \delta{}G_{0}(\vec{r}\tau,\vec{r}'0;V) &= \frac{\hbar^{2}\tau^{2}}{12m}
  \left(\frac{m}{2\pi\uI\tau}\right)^{3/2}\uE^{-\uI{}V(\vec{r})\tau}
  \left[\nabla^{2}V-\frac{\uI\tau}{2}(\vec{\nabla}V)^{2}\right] \nonumber \\
  &= \frac{\hbar^{2}\tau^{2}}{12m}
  \int\!\!\frac{\uD^{3}\vec{p}}{(2\pi\hbar)^3}
  \;\uE^{-\uI[(\vec{p}^{2}\tau/2m)+V(\vec{r})\tau]}\left[\nabla^{2}V
  -\frac{\uI\tau}{2}(\vec{\nabla}V)^{2}\right].
\end{align}
Upon integrating over $\tau$ in (\ref{Eqn01}) by parts, we obtain
for
\begin{equation*}
  \frac{2}{2\pi\uI}\int_{-\infty}^{\infty}\!\frac{\uD\tau}{\tau-\uI\varepsilon}
  \;\uI\frac{\partial}{\partial\tau}\delta{}G_{0}(\vec{r}\tau,\vec{r}'0;V)
\end{equation*}
the remarkably simple expression
\begin{equation}\label{Eqn09}
  \frac{\hbar^2}{6m}\left[\nabla^{2}V+\frac{(\vec{\nabla}V)^{2}}{2}
  \frac{\uD}{\uD{}V}\right]\int\!\!\frac{\uD^{3}\vec{p}}{(2\pi\hbar)^3}\;
  \delta\!\left(\frac{\vec{p}^{2}}{2m}+V(\vec{r})\right)
\end{equation}
involving only \emph{one} (!) derivative with respect to $V$\@. \
The latter expression is readily integrated to yield, after
straightforward rearrangements of terms,
\begin{equation}\label{Eqn10}
  \frac{1}{24\pi^{2}\hbar}\left[(\nabla^{2}V)(-2mV)^{1/2}
  -\frac{1}{3m}\vec{\nabla}\Cdot\left(\vec{\nabla}(-2mV)^{3/2}\right)\right].
\end{equation}
The second expression in (\ref{Eqn10}) gives a zero surface
contribution to (\ref{Eqn01}) at infinity and near the origin due
to the properties of
\begin{equation*}
  \left[\left(-2mV_{\mathrm{TF}}(\vec{r})\big.\right)^{3/2}
  -\left(-2mV_{\mathrm{C}}(\vec{r})\big.\right)^{3/2}\right]\textrm{ for }
  r\to\infty\textrm{ and }r\to{}0.
\end{equation*}
Finally we use the following important relations for the
potentials~:
\begin{equation}\label{Eqn11}
  \left[\left(-2mV_{\mathrm{TF}}(\vec{r})\big.\right)^{1/2}
  -\left(-2mV_{\mathrm{C}}(\vec{r})\big.\right)^{1/2}\right]\delta^{3}(\vec{r})
  = 0
\end{equation}
\begin{align}
  \nabla^{2}V_{\mathrm{TF}}(\vec{r}) &= 4\pi{}Ze^{2}\delta^{3}(\vec{r})
  -\frac{4}{3\pi}\left(\frac{2mZe^{2}}{r\hbar^{2}}\right)^{\!3/2}
  \big(f(x)\big)^{3/2} \label{Eqn12} \\
  \nabla^{2}V_{\mathrm{C}}(\vec{r}) &= 4\pi{}Ze^{2}\delta^{3}(\vec{r}) \label{Eqn13}
\end{align}
and the first expression in (\ref{Eqn10}) to immediately obtain
for $\delta{}E_{\mathrm{Qua}}$ in (\ref{Eqn01})
\begin{equation}\label{Eqn14}
  \delta{}E_{\mathrm{Qua}} = -\frac{4}{9\pi^2}\left(3\pi/4\right)^{2/3}
  \left(\frac{me^{4}}{\hbar^{2}}\right)Z^{5/3}\int_{0}^{\infty}\!\!\uD{}x\:
  \big(f(x)\big)^{2}
\end{equation}
accounting for the $-0.04907Z^{5/3}$ (in units of
$me^{4}/\hbar^{2}$) contribution to the ground-state energy\@. \\


\begin{thebibliography}{99}
\raggedright

\bibitem{Dreizler_1990}
Dreizler,~R.~M. and Gross,~E.~K.~U. (1990): \textit{Density
Functional Theory: An Approach to the Quantum Many-Body Problem},
Springer-Verlag, Berlin.

\bibitem{Fermi_1927}
Fermi,~E. (1927): ``Un Metodo Statistico per la Determinazione di
alcune Priorieta dell'Atome'',
\textit{Rend.~Accad.~Naz.~Lincei}~\textbf{6}, pp.~602--607.

\bibitem{Fermi_1928}
Fermi,~E. (1928): ``Ein statistische Methode zur Bestimung einiger
Eigenschaften des Atoms und ihre Anwendung auf die Theorie des
periodischen Systems der Elemente'',
\textit{Z.~Phys.}~\textbf{48}, pp.~73--79.

\bibitem{Morgan_1996}
Morgan~III,~J.~D. (1996): in \textit{Atomic, Molecular and Optical
Physics Handbook}, Drake,~G.~W.~F., ed., American Institute of
Physics, Woodbury, New York.

\bibitem{Schwinger_1981}
Schwinger,~J. (1981): ``Thomas--Fermi Model: The Second
Correction'', \textit{Phys.~Rev.}~\textbf{A24} (5),
pp.~2353--2361.

\bibitem{Thomas_1927}
Thomas,~L.~H. (1927): ``The Calculation of Atomic Fields.'',
\textit{Proc.~Camb.~Phil.~Soc.}~\textbf{23}, pp.~542--548.

\end{thebibliography}
\end{document}